\documentclass[twocolumn,showpacs,amsmath,amssymb]{revtex4}
\usepackage{graphicx}
\usepackage{bm}
\usepackage{psfig}
\begin{document}

\title{Tuning of the spin-orbit interaction in two-dimensional GaAs holes via strain}

\author{B.~Habib, J.~Shabani, E.~P. De~Poortere, M.~Shayegan}

\affiliation{Department of Electrical Engineering, Princeton
University, Princeton, NJ 08544, USA}

\author{R.~Winkler}
\affiliation {Department of Physics, Northern Illinois University,
De Kalb, IL, 60115, USA}

\date{\today}

\begin{abstract}
We report direct measurements of the spin-orbit interaction induced
spin-splitting in a modulation-doped GaAs two-dimensional hole
system as a function of anisotropic, in-plane strain. The change in
spin-subband densities reveals a remarkably strong dependence of the
spin-splitting on strain, with up to about $20$\% enhancement of the
splitting upon the application of only about $2 \times 10^{-4}$
strain. The results are in very good agreement with our numerical
calculations of the strain-induced spin-splitting.
\end{abstract}

\pacs{Valid PACS appear here}
\maketitle


Manipulation of the spin-orbit coupling in materials that lack
inversion symmetry is considered the basis for novel spintronic
devices \cite{DattaAPL90, NittaAPL99, Reviews}. In two dimensions,
these devices utilize the fact that the inversion asymmetry of the
confining potential can be tuned with a perpendicular electric field
applied via external front- and back-gate biases \cite{RashbaJP84,
NittaPRL97, LuPRL98, PapadakisSCI99, PapadakisPHE01, Winkler03}.
This structural inversion asymmetry, along with the bulk inversion
asymmetry of the zinc-blende structure, leads to a lifting of the
spin degeneracy of the energy bands even in the absence of an
applied magnetic field. The energy bands at finite wave vectors are
split into two spin subbands with different energy surfaces,
populations, and effective masses. It is the manipulation of this
so-called zero-field spin-splitting that forms the underlying
principal of many spintronic devices. In addition, the spin-orbit
interaction-induced spin-splitting is of interest in studying
fundamental phenomena such as Berry's phase \cite{MorpurgoPRL98,
YauPRL02} and the spin Hall effect \cite{EngelsCM06}.

There have been recent reports of utilizing strain for tuning the
spin-orbit interaction and the resulting spin-splitting
\cite{KatoNat04, SihPRB06, CrookerPRL05, BeckCM06}. The studies have
focused on magneto-optical (Faraday/Kerr rotation) measurements in
epitaxially grown but bulk-doped GaAs and InGaAs electron systems.
Here we present strain-induced spin-splitting results for a
high-mobility, modulation-doped GaAs \emph{two-dimensional hole}
system (2DHS). We utilize a simple but powerful technique to
continuously apply quantitatively measurable in-plane strain
\emph{in-situ} \cite{ShayeganAPL03}, and make magneto-transport
measurements which directly probe the densities of the
spin-subbands. We observe a significant change in spin-splitting as
a function of strain. The experimental data agree very well with our
accurate numerical calculations of the spin-splitting which take the
spin-orbit interaction and strain fully into account. We show that
the mechanism that gives rise to the strain-induced spin-splitting
in hole systems is qualitatively different from the meachanism
operating in electron systems. Most importantly, the strain
enhancement of the spin-splitting for the 2D holes is about 100
times larger than for 2D electrons and, moreover, is essentially
independent of the strain direction. Combined, our results establish
the extreme sensitivity of the spin-orbit coupling in 2DHSs to
strain, and demonstrate the potential use of the 2D holes for
spintronic and related applications.

\begin{figure*}
\centering
\includegraphics{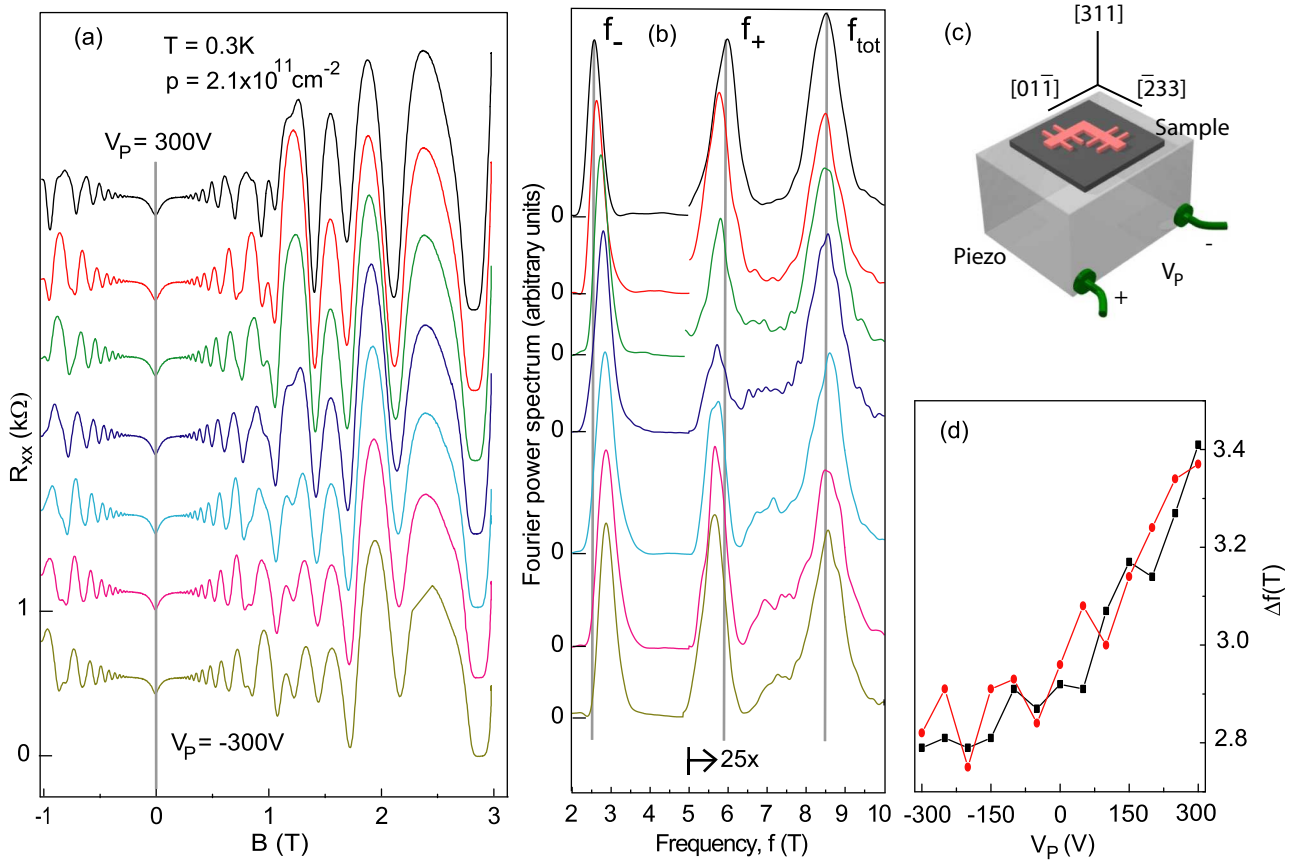}
\caption{\label{fig:fig1}(Color on-line)(a) Shubnikov-de Haas
oscillations for seven different piezo voltages in steps of $100$~V.
The traces are offset vertically for clarity. (b) Normalized Fourier
power spectra of the oscillations in the range $0.2 \leq B \leq
2~$T. The positions of the peaks $f_-$ and $f_+$ correspond to the
densities of the minority and majority spin subbands, while the peak
labelled $f_\mathrm{tot}$ gives the total 2D hole density. (c)
Experimental setup. The poling direction for the piezo is along
[$01\bar{1}$]. (d) Spin-splitting versus applied strain. The black
squares ($\Delta f = f_+ - f_-$) and the red circles ($\Delta f =
f_\mathrm{tot} - 2f_-$) are from two different methods used to
determine the spin-splitting.}
\end{figure*}

Our sample was grown on a GaAs (311)A substrate by molecular beam
epitaxy and contains a modulation-doped 2DHS confined to a
GaAs/AlGaAs heterostructure. The Al$_{0.35}$Ga$_{0.65}$As/GaAs
interface is separated from a 17~nm-thick Si-doped
Al$_{0.35}$Ga$_{0.65}$As layer (Si concentration of $4 \times
10^{18}$~cm$^{-3}$) by a 30~nm Al$_{0.35}$Ga$_{0.65}$As spacer
layer. We fabricated L-shaped Hall bar samples via photo-lithography
and used In:Zn alloyed at 440$^\circ$C for the ohmic contacts. Metal
gates were deposited on the sample's front and back to control the
2D hole density ($p$). We measured the longitudinal ($R_{xx}$) and
transverse ($R_{xy}$) magneto-resistances at $T = 0.3$~K via a
standard low frequency lock-in technique. $R_{xx}$ was measured
along [$01\bar{1}$] and [$\bar{2}33$] directions yielding, at $p =
2.1\times 10^{11}$~cm$^{-2}$, low temperature mobilities of $1.7
\times 10^5$~cm$^2$/Vs and $4.3 \times 10^5$~cm$^2$/Vs in the two
directions respectively. Here we present $R_{xx}$ data along
[$01\bar{1}$]; Measurements along [$\bar{2}33$] reveal similar
results.

We apply tunable strain to the sample by gluing it on one side of
a commercial piezoelectric (piezo) stack actuator with the
sample's [$01\bar{1}$] crystal direction aligned with the poling
direction of the piezo [Fig.~\ref{fig:fig1}(c)]
\cite{ShayeganAPL03}. When bias $V_P$ is applied to the
piezo-stack, it expands (shrinks) along the [$01\bar{1}$] for $V_P
> 0$ ($V_P < 0$) and shrinks (expands) along the [$\bar{2}33$]
direction. We have confirmed that this deformation is fully
transmitted to the sample, and using metal strain gauges glued to
the opposite side of the piezo, have measured its magnitude
\cite{ShayeganAPL03, GunawanCM06}. Based on our calibrations of
similar piezo-actuators, we estimate a strain of $3.8 \times
10^{-7} V^{-1}$ along the poling direction. In the perpendicular
direction, the strain is approximately $-0.38$ times the strain in
the poling direction \cite{ShayeganAPL03}. In this paper we
specify strain values along the poling direction; we can achieve a
strain range of about $2.3 \times 10^{-4}$ by applying $-300~ \leq
V_P \leq 300~V$ to the piezo. Finally, the back-gate on the sample
is kept at a constant voltage throughout the measurements to
shield the 2DHS from the electric field of the piezo-stack.


Figure~\ref{fig:fig1}(a) shows the low-field Shubnikov-de Haas (SdH)
oscillations for seven different values of $V_P$ from $-300$~V to
$300$~V in steps of $100$~V. The Fourier transform spectra of these
oscillations, shown in Fig.~\ref{fig:fig1}(b), exhibit three
dominant peaks at frequencies $f_-$, $f_+$, and $f_\mathrm{tot}$,
with the relation $f_\mathrm{tot} = f_+ + f_-$. The $f_\mathrm{tot}$
frequency, when multiplied by $e/h$, matches well the total 2D hole
density deduced from the Hall resistance ($e$ is the electron charge
and $h$ is the Planck constant). The two peaks at $f_-$ and $f_+$
correspond to the Fermi contours of holes in individual
spin-subbands although their positions times $e/h$ do not exactly
give the spin-subband densities \cite{Winkler03, WinklerPRL00,
KeppelerPRL02}. As we discuss below, however, this discrepancy
between $(e/h)f_{\pm}$ and the $B = 0$ spin-subband densities is
minor and $\Delta f = f_+ - f_- = f_\mathrm{tot} - 2f_-$ indeed
provides a very good measure of the spin-splitting. The vertical
gray lines in Fig.~\ref{fig:fig1}(b) clearly indicate that $\Delta
f$ increases when the piezo voltage is dialed up from $-300$~V to
$300$~V while the total hole density ($f_\mathrm{tot}$) remains
constant. Figure~\ref{fig:fig1}(d) summarizes the change in $\Delta
f$ with strain (in terms of piezo-bias) for
$p=2.1\times10^{11}$~cm$^{-2}$; $\Delta f$ determined from both
($f_+ - f_-$) and ($f_\mathrm{tot} - 2f_-$) are plotted. The results
show a significant (about 20\%) enhancement of spin-splitting with
strain.

\begin{figure}
\centering
\includegraphics{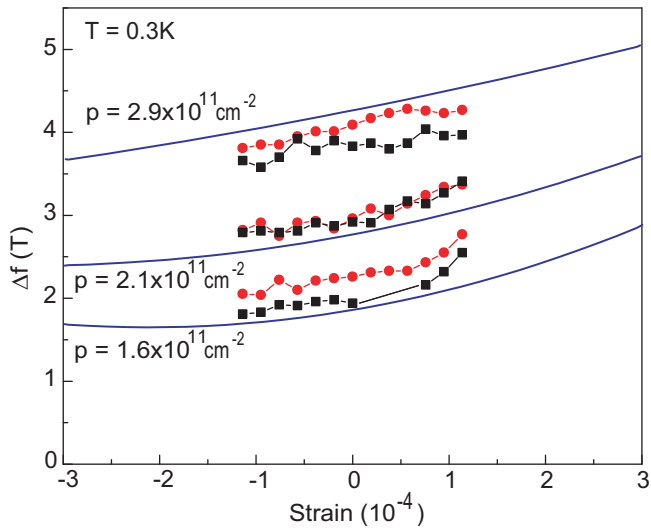}
\caption{\label{fig:fig2} (Color on-line)The black squares ($\Delta
f = f_+ - f_-$) and the red circles ($\Delta f = f_\mathrm{tot} -
2f_-$) are the experimentally measured spin-splitting. The solid
curves are $\Delta f$ determined from the calculated
magneto-oscillations.}
\end{figure}

In order to understand the data of Fig.~\ref{fig:fig1}, we performed
self-consistent calculations of the spin-splitting as a function of
strain, using the $8\times8$ Kane Hamiltonian augmented by the
strain Hamiltonian of Bir and Pikus \cite{Bir74, TrebinPRB79,
Winkler03}. This model takes into account the spin-orbit coupling
due to both the structure inversion asymmetry of the GaAs/AlGaAs
hetero-junction as well as the bulk inversion asymmetry of the
underlying zinc blende structure \cite{WinklerPRB93}. Furthermore,
it fully incorporates the strain-induced contributions to
spin-splitting. We adapted this model to the (311) orientation of
our sample by a suitable coordinate transformation. To make a direct
comparison with the experimental data, we calculated the Landau fan
chart for $B > 0$ and determined the magneto-oscillations of the
density of states at the Fermi energy \cite{WinklerPRL00,
Winkler03}. We then calculated the Fourier power spectrum of these
oscillations and obtained the frequencies $f_+$ and $f_-$ that
correspond to the majority and minority spin subbands. The
difference between these frequencies $\Delta f$ can be directly
compared to the experimentally determined $\Delta f$ data of
Fig.~\ref{fig:fig1}(d).

Figure~\ref{fig:fig2} presents our calculated $\Delta f$ (solid
curves) as a function of strain for three different 2DHS
densities. It is clear that the calculated $\Delta f$ exhibit
substantial changes with strain. In Fig.~\ref{fig:fig2} we also
show the measured $\Delta f$ values for the same three densities,
assuming that $V_P = 0$ corresponds to zero strain. There is
overall very good agreement between the calculated and measured
$\Delta f$. The agreement is particularly remarkable in view of
the fact that the calculations were performed only based on the
sample structure and density. In other words, there are no fitting
parameters used to match the results of the calculations to the
measured values of $\Delta f$.

We would like to make the following remarks about the results
presented in Fig.~\ref{fig:fig2}. First, it is known that
\cite{WinklerPRL00, KeppelerPRL02} the frequencies $f_+$ and $f_-$
are not exactly related to the spin subband densities at zero
magnetic field, $p_+$ and $p_-$, via the relation $p_\pm =
(e/h)f_\pm$, although this relation approximately holds. For
completeness, we also calculated $p_+$ and $p_-$. We find that for
the data shown in Fig.~\ref{fig:fig2}, the calculated $\Delta p$ is
only slightly larger than the calculated $(e/h)\Delta f$, by at most
10\%. This means that the results presented in Fig.~\ref{fig:fig2}
closely represent the spin-splitting at zero magnetic field as well.
Second, although there are no fitting parameters in comparing the
experimental and calculated $\Delta f$, we do have an experimental
uncertainty regarding the absolute value of strain. In our
experiments, we know the relative changes in values of strain
accurately, but we do not know the piezo-bias corresponding to
zero-strain. Thanks to a mismatch between the thermal expansion
coefficients of GaAs and the piezo-stack, at low temperatures the
sample can be under finite strain even at $V_P = 0$. This residual
strain is cooldown-dependent and unfortunately we do not know its
magnitude for the data of Fig.~\ref{fig:fig2} which were measured
during a single cooldown. Based on our experience with cooldowns of
samples glued to similar piezo-stacks, we expect a residual strain
up to about $\pm3 \times 10^{-4}$ \cite{GunawanCM06,
ShkolnikovAPL04}. If we assume that there is indeed a finite
residual strain in our experiment, and shift the experimental data
points in Fig.~\ref{fig:fig2} horizontally to the right by about $1
\times 10^{-4}$, we would find better agreement between the
calculated and measured $\Delta f$ for the $p = 1.6$ and $2.1 \times
10^{11}$~cm$^{-2}$ data; the agreement for $p = 2.9 \times
10^{11}$~cm$^{-2}$ data, however, worsens. We emphasize that despite
this uncertainty regarding the exact value of the zero-strain
condition in our study, the overall agreement of the experimental
and calculated spin-splitting, including its strain dependence, is
remarkable.

Our analysis reveals that the mechanism leading to the strain
dependence of spin-splitting in 2DHSs is very different from the
mechanism responsible for the strain-dependent spin-splitting in
bulk-like electron systems studied previously \cite{KatoNat04,
SihPRB06, CrookerPRL05, BeckCM06}. In the latter case, strain has
generally only a weak effect on the energy dispersion, and the
spin-splitting can be traced back to a small deformation potential,
often denoted as $C_2$ that couples electron and hole states in a
spin-dependent manner \cite{TrebinPRB79}. Furthermore, the strain
dependence is highly anisotropic. No spin-splitting occurs for
strain along the crystallographic [100] direction
\cite{CrookerPRL05, BeckCM06}. In 2DHSs, on the other hand, the
piezo-induced strain has a two-fold effect. First it changes the
heavy-hole light-hole (HH-LH) energy splitting. Since spin-splitting
in 2DHSs is known to compete with the HH-LH splitting
\cite{Winkler03}, this provides a direct way to tune the
spin-splitting. Second, the strain changes the functional form of
the spin-splitting of 2DHSs. While in the absence of strain the
spin-splitting of 2DHSs is cubic in the wave vector $k$
\cite{Winkler03}, strain gives rise to a significant spin-splitting
linear in $k$. The deformation potentials relevant for these effects
(often denoted as $D_u$ and $D_u'$ \cite{TrebinPRB79}) are much more
important than $C_2$ which is the reason why the strain dependence
of spin-splitting is much more pronounced in hole systems as
compared to electron systems. Also, the strain-induced
spin-splitting in 2DHSs depends only weakly on the direction of the
in-plane strain. The effect of the spin-dependent deformation
potential $C_2$ on the spin-splitting of hole states is much smaller
than the effects discussed here.

For a more quantitative comparison, we deduce the spin splitting in
a 2D GaAs electron system at a density of $2.1\times
10^{11}$cm$^{-2}$ by using Eq. (6.18) in Ref. \cite{Winkler03}. The
change in spin-subband densities for this system for an applied
strain of $1\times 10^{-4}$ is $8.2\times10^{7}$~cm$^{-2}$. From
Fig. \ref{fig:fig2} the corresponding change in spin-subband
densities for our sample is $8.4\times10^{9}$~cm$^{-2}$. This is
\emph{100 times} larger than the 2D electron system
\cite{Footnote2DEG}.

We close by highlighting a potential application of our findings.
Our results reveal a surprisingly large change in spin-splitting in
GaAs 2D holes for rather small values of applied strain. This tuning
of the spin-splitting can be employed to demonstrate various
spintronic and/or spin-interference effects in devices, such as
Aharonov-Bohm type ring structures, made in this system. In the spin
interference device proposed by Nitta \emph{et al.}
\cite{NittaAPL99}, \emph{e.g.}, the conductance through a ring of
radius \emph{a} is expected to oscillate with a period of $\delta
(\Delta k) \pi a$, where $\delta (\Delta k)$ denotes the change in
$\Delta k$, defined as the difference of the Fermi wave vectors of
the two spin-subbands. In our 2DHS sample, at $p =$ $2.1 \times
10^{11}$~cm$^{-2}$, the strain-induced change in $\Delta k$ is $\sim
0.9 \times 10^7$~m$^{-1}$ \cite{Footnotedk}. Hence for a ring of
radius $220$~nm, the conductance through the ring would go through
one period of oscillation. For such measurements, tuning the
spin-orbit interaction via strain, rather than perpendicular
electric field (gate bias), may prove advantageous: the 2D hole
density remains fixed as a function of strain, thus simplifying the
experimental measurements and their interpretation \cite{Footnote1}.

After the completion of this work we learned of related results by
Kolokolov \emph{et al.} \cite{KolokolovPRB99} who studied the strain
dependence of the spin-splitting in GaAs (100) 2DHSs. Their
experimental data, however, are only in qualitative agreement with
their calculations.

We thank the DOE, ARO, NSF and the Alexander von Humboldt Foundation
for support.


\begin{thebibliography}{99}
\bibitem{Footnotedk}
$\Delta k$ is deduced from the spin-subband densities by assuming an
isotropic dispersion giving $k_\pm = \sqrt{4\pi p_\pm}$. This
simplifying assumption helps in understanding the ring experiment
qualitatively. A more detailed analysis would require taking the
anisotrpy of the HH band into account.
\bibitem{Footnote1}
Tuning the spin-splitting via gate bias has the often undesired
result of changing the total 2D density and therefore the number
of conducting channels in the ring.
\end{thebibliography}
\end{document}